# Computational Analysis of Modified Blom's Scheme

Suraj Sukumar

**Abstract**: To achieve security in wireless sensor networks, it is important to be able to encrypt and authenticate messages sent among sensor nodes. Due to resource constraints, achieving such key agreement in wireless sensor networks is nontrivial. Blom's scheme is a symmetric key exchange protocol used in cryptography. But the concerns with Blom's scheme are the complexity involved in computation as well as memory usage. In this paper, we propose a new key pre-distribution scheme by modifying Blom's scheme which reduces the computational complexity as well as memory usage.

## Introduction

Key exchange schemes for sensor networks are a subject of continuing interest and one of the techniques used includes Blom's scheme and its variants [1]-[6]. The key distribution method proposed by Blom allows any pair of nodes in a network to be able to find a pair wise secret key [7]. This network consists of N nodes termed as N users and a collusion of less than $\lambda+1$ users cannot reveal the keys which are held by other users. $\lambda$ which is the threshold is treated as a security parameter i.e., the larger the threshold $\lambda$ value the more secure the network. This threshold property of Blom's scheme is a desirable feature because an adversary needs to attack a significant fraction of the network in order to achieve high payoff. Another usage of $\lambda$ is that it helps in determining the amount of memory needed to store the key information. Higher the $\lambda$ value leads to higher memory usage. The network is said to be secure as long as no more than $\lambda$ nodes are compromised. This is called the $\lambda$- secure property.

The working of the Blom's pre-distribution scheme is described below. For context related to secure networks, error-correction, and implicit security see [8]-[11]. Since we know that the Blom's scheme was not developed for wireless sensor networks, modifications have been done to the Blom's scheme to make it suitable for the wireless sensor networks.

**Blom's Scheme**

In the predevelopment phase, over a finite set of field GF(q), the base station constructs a matrix G of size $(\lambda +1) \times N$. where N is the number of nodes, i.e., the size of the network and $\lambda$ is the secure parameter. G is called the public matrix i.e., all the information of G is public information, which means every sensor node in the network can access the information of G and the same goes with the advisories. Next the base station, over a finite field GF(q), creates a symmetry matrix called D of size $(\lambda + 1) \times (\lambda + 1)$. Now the base station computes $(D.G)^T$ which results in a matrix called A of size $N \times (\lambda + 1)$, where $(D.G)^T$ is the transpose of D.G. Matrix D needs to be kept secret, and should not be disclosed to adversaries or any sensor node (although, as will be discussed later, one row of $(D.G)^T$ will be disclosed to each sensor node). Because D is symmetric, it is easy to see:

$$A.G = (D.G)^T.G = G^T.D^T.G = G^T.D.G$$
$$= (A.G)^T;$$

From the above equation we can see that $A.G = (A.G)^T$ which means A.G is a symmetric matrix. Let us store the resultant of A.G in a new matrix called K. Since we know A.G is a symmetric matrix, we can say an element in K located at ith row and jth column is equal to an element in K at jth row and ith column i.e., $K_{ij} = K_{ji}$. Hence we use $K_{ij}$ (or $K_{ji}$) as the pairwise key between node i and node j. Fig. 1 illustrates how the pair wise key $K_{ij} = K_{ji}$ is generated. To carry out the above computation, nodes i and j should be able to compute $K_{ij}$ and $K_{ji}$, respectively. This can be easily achieved using the following key pre-distribution scheme, for k = 1. . . N:

1. store the $k^{th}$ row of matrix A at node k, and
2. store the $k^{th}$ column of matrix G at node k.

Therefore, when nodes i and j need to find the pair wise key between them, they first exchange their columns of G, and then they can compute $K_{ij}$ and $K_{ji}$, respectively, using their private rows of A. Because G is public information, its columns can be transmitted in plaintext. It has been proved in [4] that the above scheme is λ-secure if any λ + 1 columns of G are linearly independent. This λ-secure property guarantees that no nodes other than i and j can compute $K_{ij}$ or $K_{ji}$ if no more than λ nodes are compromised.

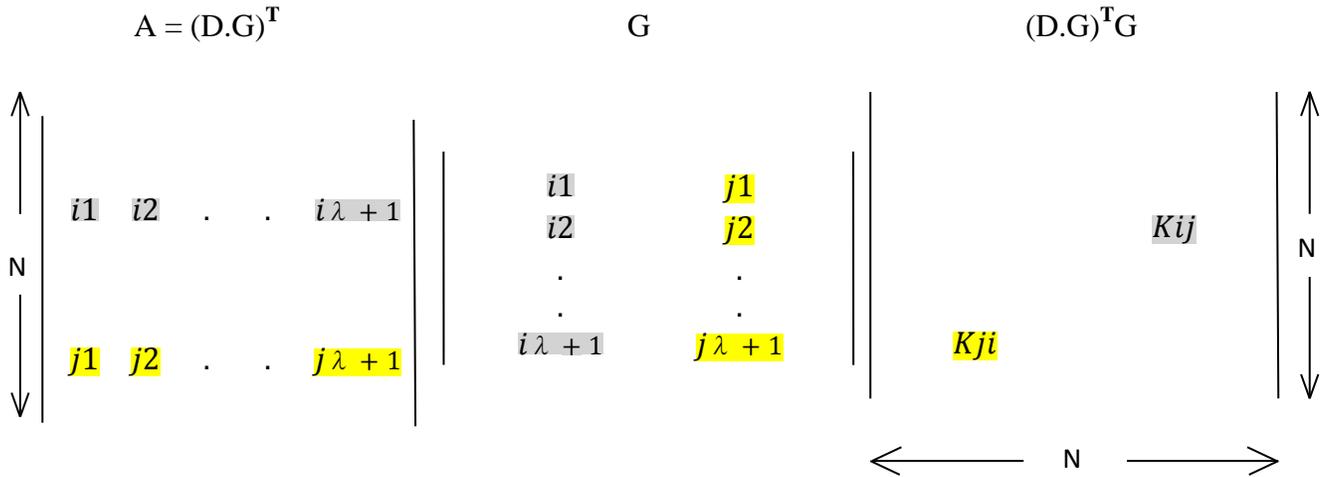

Figure 1: Generating Keys in Blom's Scheme

Below is how a Matrix G looks like. A Vandermonde matrix is a matrix with the terms of a geometric progression in each row. And since we would like to achieve λ-secure property, any (λ+1) columns of G must be linearly independent in order. Since every element in the finite field GF(q) represents a pair-wise key, suppose the length of pair-wise key is 128, then q should be choosen in such a way that it is the smallest prime number that is larger than $2^{128}$. In a given finite field GF(q) where N<q let a be a primitive element of GF(q). So now each non zero element of GF(q) can be represented by some power of s. namely $s^i$ for some $0 < i \leq q - 1$. A feasible G can be designed as follows [7]:

$$G = \begin{bmatrix} 1 & 1 & 1 & \cdots & 1 \\ s & s^2 & s^3 & \cdots & s^N \\ s^2 & (s^2)^2 & (s^3)^2 & \cdots & (s^N)^2 \\ & & \vdots & & \\ s^\lambda & (s^2)^\lambda & (s^3)^\lambda & \cdots & (s^N)^\lambda \end{bmatrix}$$

Figure 2: Public Matrix example

We know that $s^i \neq s^j$ if $i \neq j$ as this is the property of primitive elements. As we have already stated a Vandermonde matrix is a matrix with the terms of a geometric progression in each row, we can say that $\lambda + 1$ columns of G are linearly independent when s, $s^2$, $s^3$, ..., $s^N$ are all distinct. In practice a primitive element s of GF(q) can generate matrix G. Hence any node when given a seed can generate the column, i.e., when we store the $k^{th}$ column of G at node k, we only need to store the seed $s^k$ at this node.

**Modified Blom's Scheme**

In this article, we make use of the Bloms scheme. As stated earlier the Bloms scheme makes use of Vandermonde matrix which is a public matrix and this matrix is responsible for all computations in generating keys and since it's a public matrix, it could be known even to the eavesdroppers. We choose all the values or elements of this matrix to be distinct so that any $\lambda + 1$ columns of G are linearly independent i.e., to generate unique keys. But when the value of $\lambda$ increases to larger values, then the number of rows of the public matrix increase and this in turn results in greater value in the columns because the column values increases in a geometric series. We know in wireless sensor networks, sensor nodes contains limited memory and energy, where in Bloms scheme [4] for any two nodes to generate a common key, each node should store column of public matrix and row of the calculated secret matrix and this would be difficult for sensor networks to store both the row and column in the memory for a large network.

To reduce the computation and memory overhead in Blom's scheme, instead of using Vandermonde matrix [12] we propose the use an Adjacency Matrix as the public matrix. As, the Adjacency matrix is a square matrix with 1s and -0s, it reduces of complexity of calculating values for all the elements corresponding to the columns in Vandermonde matrix. This Adjacency matrix is formed in such a way that all nodes that are neighbors of a particular node are filled with 1s and remaining with q-1(since public matrix cannot contain 0s).

Another advantage of using Adjacency matrix is it reduces the cost of saving the columns in the memory of sensor because any node can easily generate Adjacency matrix of known size. Similar to Blom's scheme the operation which are to be performed to generate the keys will depend on the prime number i.e., the number which depends on the desired key length. A normal Adjacency matrix would look like the one in figure 3 but our modified Adjacency matrix would look like figure 4.

$$\begin{vmatrix} 1 & 1 & 0 & 1 \\ 1 & 0 & 0 & 0 \\ 0 & 0 & 1 & 1 \\ 1 & 0 & 1 & 0 \end{vmatrix}$$

Figure 3: Adjacency Matrix

$$\begin{vmatrix} 1 & 1 & 28 & 1 \\ 1 & 28 & 28 & 28 \\ 28 & 28 & 1 & 1 \\ 1 & 28 & 1 & 28 \end{vmatrix}$$

Figure 4: Changed modified Adjacency Matrix for Modulo 29

The original binary form Adjacency Matrix is depicted in Figure 3. The only change that is made to the Adjacency Matrix is all zeros are replaced with the q-1(prime number -1). We can see from Figure 4 that the Adjacency Matrix consists of only two different numbers one's and q-1(prime number -1), hence all the further calculations will be very simple. Key generation technique is similar to that used in the Blom's scheme [7]. Following are the steps involved in calculating the key.

> **Step 1: Generating a G matrix.** We first select a primitive element from a finite field GF(q), where q is larger than the desired key length (and also q > N), and then construct adjacency matrix G of size N ×N as discussed in the previous section depending on the node neighbors. Then depending on the $\lambda$ value first $\lambda$ rows along with N columns are selected as the public matrix. Let G (j) represent the $j^{th}$ column of G, in Blom's scheme G (j) is provided to node j but in our case each node knows its neighbors and therefore the memory usage for storing G (j) at a node is not required.

**Step 2: Generating key spaces.** The Central Authority generates $\omega$ random, symmetric matrices $D_1, \ldots, D_\omega$ of size $(\lambda+1)\times(\lambda+1)$. We then compute the matrix $A_i = (D_i \cdot G)$ T. Let $A_i(j)$ represent the $j^{th}$ row of $A_i$.

**Step 3: Computing Secret Key.** The central authority stores each row of the matrix A in the node memory with corresponding index. Now if node i want to communicate with node j then node i multiplies the row $A_i$ with column $G_j$ and the result will be the secret key.

## Implementation

The following example shows the working of the modified Blom's scheme using adjacency matrix. Let the number of nodes in the network be 6 (N=6), secure parameter $\lambda = 3$ and prime number q=29 which says if no more than 3 nodes in the network are compromised it is not possible to find the keys of other users.

Consider the node structure below:

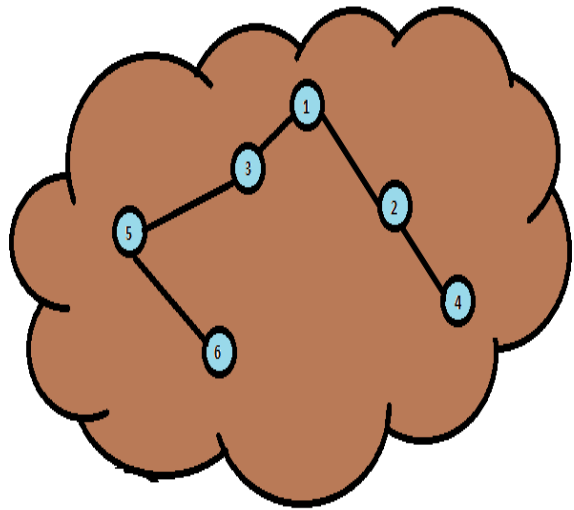

Figure 5: Node Structure Example

The Adjacencymatrix for the Figure 4 is shown below

**Modified Adjacencymatrix**

$$\begin{vmatrix} 28 & 1 & 1 & 28 & 28 & 28 \\ 1 & 28 & 28 & 1 & 28 & 28 \\ 1 & 28 & 28 & 28 & 1 & 28 \\ 28 & 1 & 28 & 28 & 28 & 28 \\ 28 & 28 & 1 & 28 & 28 & 1 \\ 28 & 28 & 28 & 28 & 1 & 28 \end{vmatrix}$$

**Public matrix (G)**

$$\begin{vmatrix} 28 & 1 & 1 & 28 & 28 & 28 \\ 1 & 28 & 28 & 1 & 28 & 28 \\ 1 & 28 & 28 & 28 & 1 & 28 \\ 28 & 1 & 28 & 28 & 28 & 28 \end{vmatrix}$$

**Secret Symmetric Matrix (D)**

$$\begin{vmatrix} 3 & 5 & 2 & 7 \\ 5 & 6 & 9 & 1 \\ 2 & 9 & 3 & 5 \\ 7 & 1 & 5 & 4 \end{vmatrix}$$

**A= (D.G)$^T$ mod 29**

$$\begin{vmatrix} 26 & 9 & 5 & 24 \\ 3 & 20 & 24 & 5 \\ 18 & 18 & 14 & 26 \\ 22 & 20 & 28 & 14 \\ 16 & 26 & 16 & 22 \\ 12 & 8 & 10 & 12 \end{vmatrix}$$

Once matrix A is calculated each sensor node memory is filled with unique row chosen from matrix A corresponding to the same index. These rows represent the private keys of each node.

**Key Generation**

Suppose let's consider two nodes, node 2 and node 5 want to communicate with each other. In order to obtain the shared secret key, node 2 will multiply its private row which is provided to it by the Central Authority from matrix A which is $A_{(2)}$ with the public column of node 5 which is column 5 of G ($G_{(5)}$).

Similarly node 5 multiplies its private row $A_{(5)}$ with public column of node 2 $G_{(2)}$. In this way both node 2 and node 5 generate secret shared key for them to communicate.

$$K_{2,5} = A_2.G_5 = \begin{vmatrix} 3 & 20 & 24 & 5 \end{vmatrix} \begin{vmatrix} 28 \\ 28 \\ 1 \\ 28 \end{vmatrix} = 808 \bmod 29 = 25$$

$$K_{5,2} = A_5.G_2 = \begin{vmatrix} 16 & 26 & 16 & 22 \end{vmatrix} \begin{vmatrix} 1 \\ 28 \\ 28 \\ 1 \end{vmatrix} = 1214 \bmod 29 = 25$$

We observe that both nodes generate a common key and further communication between them will make use of the pair-wise key.

In general matrix K can be represented as shown below and we can notice that the symmetric nature gives the same key for any pair of nodes such that $K_{ij} = K_{ji}$.

K= A.P

$$K = \begin{vmatrix} 1414 & 442 & 1090 & 1549 & 1657 & 1792 \\ 268 & 1240 & 1375 & 916 & 808 & 1456 \\ 1264 & 940 & 1642 & 1642 & 1750 & 2128 \\ 1056 & 1380 & 1758 & 1812 & 1596 & 2352 \\ 1106 & 1214 & 1808 & 1538 & 1808 & 2240 \\ 690 & 528 & 852 & 960 & 906 & 1176 \end{vmatrix}$$

## Simulation

In this paper, the effort required to make the computations is analyzed. The computational effort here is calculated as the number of minute operations required to complete one Standard Arithmetic Operation. In Blom's scheme the Standard Arithmetic Operation is Multiplication. When we multiply two numbers, say at least one number is a two or higher digit number, manually to get the final result we make a number of multiplications and additions (internally). We have performed analysis between the Bloms Scheme and the proposed scheme based on this computational effort.

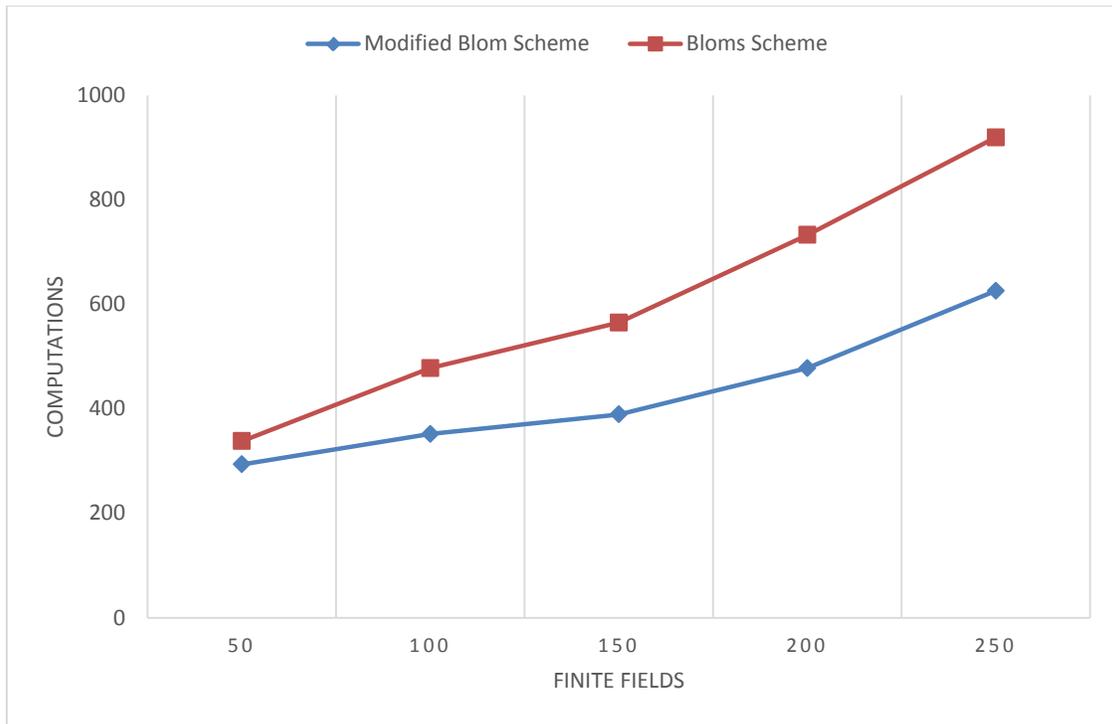

Figure 6: Computation Effort for 6 node structure

Simulations are carried out with the network size of 6 nodes and 8 nodes under a finite field ranging from 0-50, 0-100, 0-150, 0-200, 0-250, 0-300, 0-350 considering some random node structure. Both these networks have been tested with different λ values, 3 and 6 respectively. This simulation is performed 10 times and the average results are taken in consideration.

For the Blom's scheme we have generated the public and secret matrix by generating random numbers within the finite field. The same secret matrix is used for the Modified Blom's Scheme and the public matrix is formed by using the proposed scheme where the random number is also generated randomly under the finite field. The results have shown the computational effort required by the Modified Blom's Scheme is less when compared to that of the Blom's Scheme.

For the six nodes structure the difference between Bloms scheme computations and Modified Bloms scheme computations for each finite field is shown below:

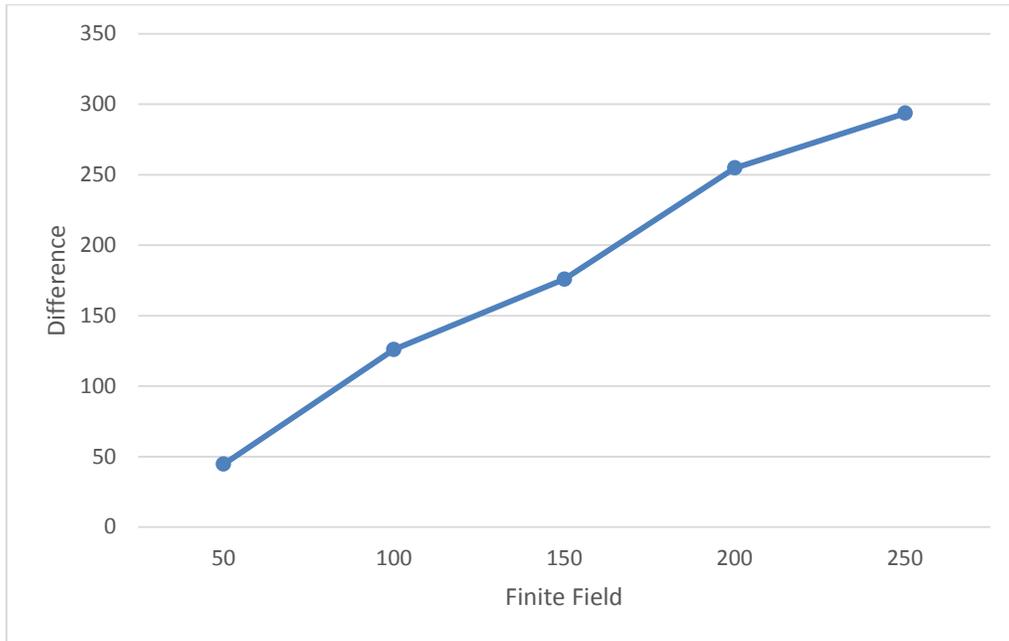

Figure 7: Difference between Bloms scheme and Modified Blom scheme

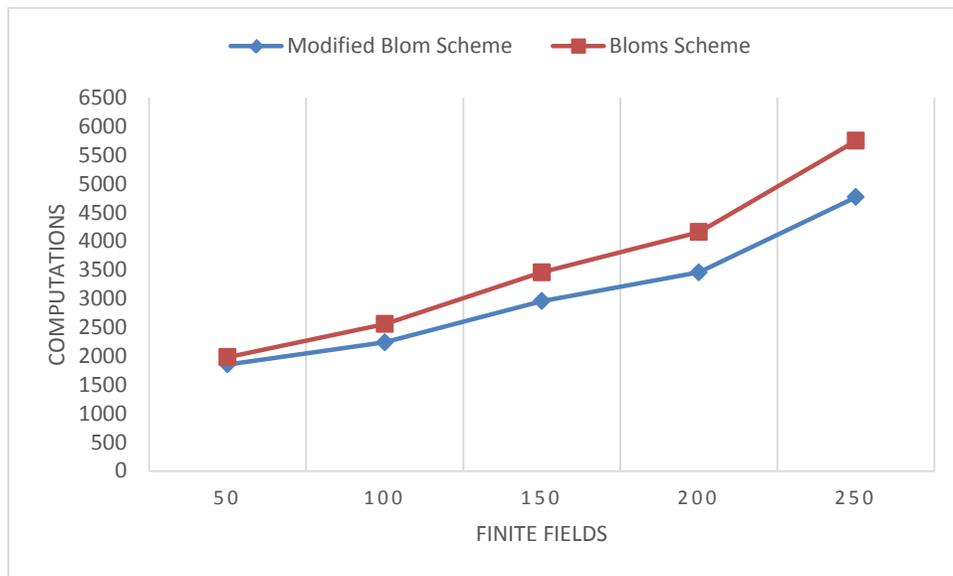

Figure 8: Computation Effort for 8 node structure

For the eight nodes structure the difference between Bloms scheme computations and Modified Bloms scheme computations for each finite field is shown below:

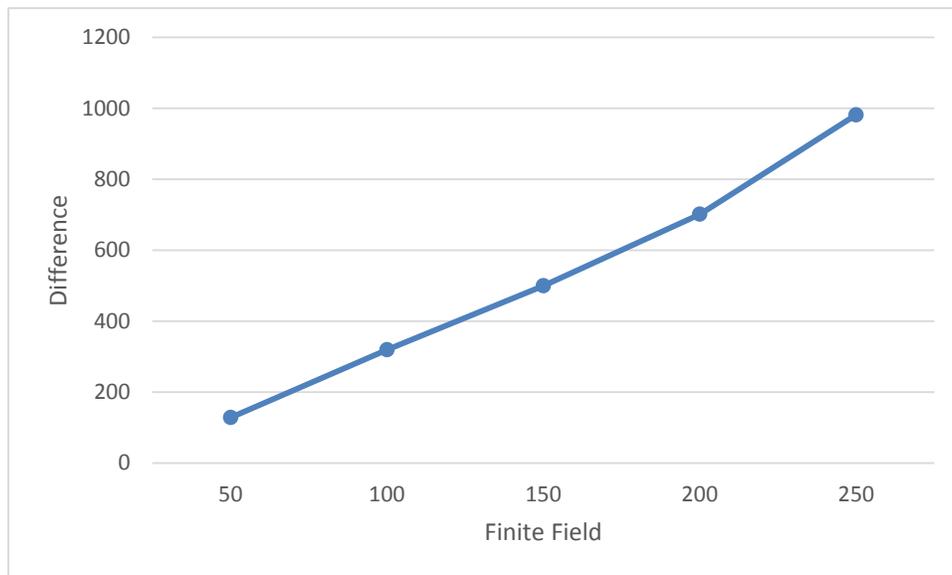

Figure 9: Difference between Bloms scheme and Modified Blom scheme

From both the results we can observe that the effort required in case of Modified Blom's scheme is less when compared to that of Blom's scheme and also we can observe that as we tend to larger finite field, the gap between the computational effort is increasing.

## Conclusion

Blom's scheme is a symmetric key exchange protocol used in cryptography. But the concerns with Blom's scheme are the complexity involved in computation as well as memory usage. This paper presents a new key pre-distribution scheme by modifying Blom's scheme which reduces the computational complexity as well as memory usage. The analysis is substantiation by simulations carried on two networks with different node size and different finite fields.